# Novel multifunctional $^{90}$Y-labelled albumin magnetic microspheres for cancer therapy

S. Vranješ-Đurić [1,5*], M. Radović [1,5], N. Nikolić [1], D. Janković [1], G. F. Goya [2,3], T. E. Torres [2,3], M. P. Calatayud [2], I. J. Bruvera [2], M. R. Ibarra [2,3], V. Spasojević [1], B. Jancar [4] and B. Antić [1]

[1] Institute of Nuclear Sciences "Vinča", University of Belgrade, P.O. Box 522, 11001 Belgrade, Serbia

[2] Instituto de Nanociencia de Aragón (INA), University of Zaragoza, Mariano Esquillor s/n, 50018- Zaragoza, Spain

[3] Departamento de Física de la Materia Condensada, Facultad de Ciencias, University of Zaragoza, Spain 50009

[4] Jožef Štefan Institute, Jamova 39, 1000 Ljubljana, Slovenia

[*] Corresponding author: Tel./fax: +381 11 80 66 436
E-mail address: sanjav@vinca.rs
[5] These authors equally contributed to the work

**Abstract**

We present *in vitro* and *in vivo* studies of yttrium-90 ($^{90}$Y)-labelled human serum albumin magnetic microspheres (HSAMMS) as multifunctional agent for bimodal radionuclide-hyperthermia cancer therapy. The HSAMMS were produced using a modified emulsification-heat stabilization technique and contained 10-nm magnetite nanoparticles coated with citric acid, distributed as inhomogeneous clusters within the albumin microspheres. The average particle size of the complete HSAMMS was 20 μm, and they exhibited superparamagnetic behavior at room temperature. The stability of the $^{90}$Y-labelled HSAMMS was investigated *in vitro* (in saline and human serum) and *in vivo* by analyzing their biodistribution in normal Wistar rats. The *in vitro* experiments revealed the high stability of the labelled HSAMMS in saline and human serum after 72 h. Following the intravenous administration of the $^{90}$Y-HSAMMS in rats, 88.81% of the activity localizes in the lungs after 1 h, with 82.67% remaining after 72 h. These data on $^{90}$Y-HSAMMS provide good evidence for their potential use in bimodal radionuclide-hyperthermia cancer therapy.

## 1. Introduction

Huge progress in nanoscience and nanotechnology has led to the development of new, nanostructured materials with properties that promise breakthroughs in a vast number of potential applications. Furthermore, it has fostered the emergence of a new field, i.e., nanomedicine, with the potential for providing revolutionary approaches to the diagnosis and treatment of some fatal diseases. In particular, magnetic nanoparticles (MNPs) are finding increasing interest in a broad range of biomedical applications, including magnetic targeting, drug and gene delivery, magnetic separation, magnetic resonance imaging, and hyperthermia. MNPs have led to a novel therapy known as thermotherapy, recently approved for clinical use [1,2]. This new protocol uses magnetic nanoparticles for heating a body region through the application of alternating magnetic fields (AMF) and the energy absorption of single-domain MNPs as result of Arrhenius-Néel relaxation [3,4]. Furthermore, it has been reported that MNPs can be incorporated inside target cells to provoke a large decrease in cell viability, with no detectable increase of temperature in the cell medium [5,6,7]. A large fraction of work on hyperthermia therapy involves magnetite nanoparticles, because they are highly biocompatible, cheap, and can be made in a variety of ways. However, their use in therapy is limited, mainly due to the nonspecific targeting. Also, due to their tendency to agglomerate or to be rapidly covered with plasma proteins in the blood, they are quickly cleared by macrophages before they can reach the target cells [8]. The specific targeting can be improved either by tuning the particle size, which determines the particles' preference for a specific organ [8], or by using a large magnetic-field gradient that is capable of retaining the MNPs within the application volume around and inside the target location [6, 7].

Albumin microspheres have received a lot of attention in recent decades as efficient drug-delivery vehicles because they are easy to prepare in defined sizes and carry reactive groups (amino and carboxylic groups) on their surfaces that can be used for radionuclide, chelator, antibodies binding and/or other surface modifications. Loaded with a sufficient amount of magnetite nanoparticles these microspheres might be used in hyperthermia treatment at a localized disease site where they could be delivered, after intravenous injection, by either passive means (e.g., trapping by size) or active means (e.g., magnetic targeting). Since hyperthermia alone is ineffective in most cases of cancer it needs to be combined with other forms of cancer therapy, such as radiation therapy. Hyperthermia may make some cancer cells more sensitive to radiation or harm other cancer cells that radiation cannot damage. When hyperthermia and radiation therapy are combined, they are often given within an hour of each other [ 9].

Albumin microspheres currently have many clinical applications as carriers of diagnostic and therapeutic radionuclides. Labelled with $^{99m}$Tc they are routinely used for lung-perfusion scintigraphy [10], while by labelling with $^{188}$Re they could be applied in internal radiation therapy [11, 12, 13, 14]. Furthermore, albumin microspheres have been used as carriers of $^{67}$Ga [15], $^{68}$Ga [16], $^{177}$Lu, $^{86}$Y and $^{90}$Y [17]. In these cases, chelators were attached onto the surface of the spheres and the particles were labelled by subsequent complexation of the radionuclides. In this work yttrium-90 ($^{90}$Y) was selected for the labelling of HSAMMS as a clinically acceptable, long-range, β-emitting radionuclide with optimal nuclear-physical characteristics for use in radionuclide therapy. Yttrium-90 is a pure, high energy β-emitter (Emax$_β$ of 2.27 MeV) with a half-life of 64.1 h [18] and it can produce biological damage over a distance that is determined by the range of the radionuclide emission (mean/maximum penetration depth is 3.6/11 mm in the soft tissue) [19]. There are currently two commercially available $^{90}$Y-microsphere products – glass microspheres [20] and resin microspheres [21] – that are clinically useful for the selective internal radiation treatment (SIRT) of primary and metastatic liver cancer [22]. Besides being used in radionuclide therapy, radionuclides provide opportunities for the investigation of the physicochemical properties and biodistribution patterns of newly developed nano- and micro-particles [23].

This study reports on the synthesis of human serum albumin microspheres with encapsulated citric acid-coated magnetite nanoparticles and their effective labelling with $^{90}$Y for possible use in bimodal radionuclide and hyperthermia cancer therapy. We discuss the results from the structural and magnetic characterization as well as their efficiency as heating agents through specific power-absorption measurements. Finally, the results from a series of *in vitro* stability studies in saline and human serum as well as *in vivo* biodistribution studies of $^{90}$Y-HSAMMS in Wistar rats are presented.

## 2. Experimental details

### 2.1. Chemicals

In the experiment we used the following: ferric chloride anhydrous ($FeCl_3$ >99%, Fluka), ferrous chloride tetrahydrate ($FeCl_2$ x $4H_2O$ >99%, Sigma-Aldrich), ammonium hydroxide (25 wt%, Zorka, Serbia), citric acid (Zorka, Serbia), cottonseed oil (Beohemik, Serbia), and diethyl ether anhydrous (Baker, Holland). Purified deionized water was prepared by the Milli-Q system (Millipore Co., Billerica, MA, USA). The human serum and human serum albumin were obtained from the National Blood Transfusion Institute (Belgrade, Serbia). The $^{90}YCl_3$ was purchased from Polatom, Poland, in a no-carrier-added form (29.64 GBq $cm^{-3}$ in 0.05 M HCl).

### 2.2. Synthesis of magnetite and citric acid-coated magnetite nanoparticles

Citric acid-coated magnetite nanoparticles were prepared using a two-step approach (post-synthesis coating) [24-26] and a co-precipitation method initially developed by Massarat [27]. Aqueous solutions (20 ml each) of 0.1 M $FeCl_2·4H_2O$ and 0.2 M $FeCl_3$ were mixed together, followed by the drop-wise addition of $NH_4OH$ (25% w/w) up to pH 10 under vigorous stirring and heating at 80 °C for 30 min. The resulting magnetite ($Fe_3O_4$) precipitate, retrieved by magnetic decantation, was washed several times with deionized water until neutral pH. After the magnetite synthesis, the coating reaction was carried out by the addition of 1 ml of a 1M citric acid solution. The temperature was raised to 90 °C and under continuous stirring for 90 minutes a stable suspension of citric acid-coated magnetite particles, i.e., ferrofluid, was obtained.

### 2.3. Preparation of HSAMMS

HSAMMS were prepared using an emulsification technique, described previously [28] with slight modifications, and stabilized by heat denaturation. Cottonseed oil was used as the oil phase and a mixture of HSA (1ml 20% HSA) and the previously prepared ferrofluid (0.5 ml, 88 mg) as the water phase. The HSA-ferrofluid mixture was added slowly (during 10 min) dropwise with a syringe using a 21-gauge needle into the cottonseed oil at room temperature and vigorously stirred for 30 min. Afterwards the temperature was increased to 100 °C and the mixture was stirred for another 30 min. This resulted in the rapid evaporation of the existing water and the albumin's irreversible denaturation, causing the formation of solid microspheres. The HSAMMS were filtered on Whatman No.5 filter paper, washed several times with ether to remove any adherent oil and vacuum dried in a desiccator overnight. The final HSAMMS were stored at 4 °C until use. The uniformity of the size of the microspheres was checked with light microscopy.

### 2.4. Experimental methods for characterization of the samples

An integrated study of the structure of the microspheres was made by combining complementary techniques, such as XRD, FESEM, SEM/FIB dual beam and HRTEM. FTIR spectroscopic analyses were carried out at room temperature using a Nicolet 380 spectrophotometer in the spectral range 4000–400 $cm^{-1}$, with a resolution of 4 $cm^{-1}$. X-ray powder diffraction (XRPD) data were collected on a Philips PW1710 diffractometer. The data were collected in the angular range 10-70 ° (2θ) with a step size of 0.06º and a counting time of 50 s per step. The thermogravimetric (TGA) analyses were performed simultaneously (30-1000 °C range) on a SDT Q600 TGA/DSC instrument (TA Instruments). The heating rates were 20 °C min$^{-1}$ and the sample mass was less than 10 mg. The furnace atmosphere consisted of air at a flow rate of 100 cm$^3$ min$^{-1}$.

Field-emission scanning electron microscopy (FESEM) was used in order to determine the morphology, size and size distribution of the HSAMMS. A Jeol JSM-7600F with a Schottky-type emitter was used at an accelerating voltage of 1.5 kV. The internal structure of the HSAMMS particles was analyzed using a SEM/FIB dual-beam device (NovaTM 200 NanoLab, FEI Company). The two beams were configured to allow the SEM to provide high-resolution imaging of the surface of a cross-section cut into the sample with the FIB.

High-resolution transmission electron microscopy (TEM) measurements were carried out using a Jeol JEM 2100 HR electron microscope operating at 200 kV. The samples were prepared by ultra-sonification in ethanol and deposited on a conventional carbon-covered copper HRTEM grid. After drying, the samples were examined by HRTEM.

Hysteresis loops were measured at 300 K in the zero-field-cooled (ZFC) regime using the Quantum Design SQUID-based magnetometer MPMS XL-5.

The microspheres were submitted to alternating magnetic fields (AMF) using a home-made applicator, consisting of a resonant LC tank working at close to the resonant frequency. The specific power absorption (SPA) of the samples ($f$ = 355 kHz, H = 47 kA/m) was measured inside a thermally insulated Dewar. Temperature data were taken using a fiber-optic temperature probe immune to the RF environment. Before each experiment, the temperature evolution was measured for 5-10 min, with the RF source turned off, in order to establish a T-baseline.

### 2.5. Labelling of HSAMMS with yttrium-90

For the radiolabelling of HSAMMS, to a suspension of 10 mg of HSAMMS in 1 ml at pH 4.2, 5 µl of $^{90}YCl_3$ solution containing approximately 370 MBq was added. The suspension was shaken and heated at 80 °C for 30 min. Labelled HSAMMS were then recovered by magnetic decantation and washed once with water to eliminate the unbound $^{90}Y^{3+}$. The final $^{90}Y$-HSAMMS were re-suspended in 2ml of 0.9 % NaCl solution (saline) and used in the subsequent studies.

The labelling yield was determined as the ratio of the measured HSAMMS-associated activity obtained after magnetic separation and the known activity used for the labelling. The radiolabelled particles were magnetically precipitated, but because of the risk of withdrawing labelled HSAMMS, the supernatant was not removed completely, so the activity was measured separately in a 0.1 ml aliquot of the supernatant and the remaining sample (radiolabelled HSAMMS and 1.9 ml supernatant). The acquired radioactivity was calculated as follows: the activity of the remainder – the activity of the 0.1 ml supernatant x1.9. Another method used for the determination of the unbound $^{90}Y^{3+}$ present in the supernatant was instant thin-layer chromatography (ITLC) on silica-gel-impregnated glass-fiber sheets (SG) in normal saline, since it is a well-accepted and uncomplicated process to examine the radiopharmaceutical quality in nuclear medicine. In this condition, soluble free $^{90}Y^{3+}$ moved to the solvent front ($R_f$ =0.9-1.0), while the $^{90}Y$- HSAMMS remained at the origin ($R_f$ =0.0-0.1).

Since yttrium-90 is a pure β-emitter it can only be detected by "bremsstrahlung" [29, 30] measured in a cross-calibrated well counter Capintec CRC-15 beta counting calibrator (Ramsey, NJ USA) and well-type NaI(Tl) gamma counter (Wallac Compu Gamma Counter LKB,

Finland). The results are reported as the mean of at least three measurements ± standard deviation.

## 2.6. *In vitro* stability of $^{90}$Y-labelled HSAMMS

*In vitro* stability studies were performed by incubating 2 mg of $^{90}$Y-HSAMMS in 2 ml of either saline or human serum at 37 °C. At various times after the preparation the radiolabelled particles were magnetically precipitated and the activity was measured separately in a 0.1 ml aliquot of the supernatant and the remaining sample (radiolabelled HSAMMS and 1.9 ml supernatant). The HSAMMS bound radioactivity was calculated as follows: the activity of the remainder – the activity of the 0.1 ml supernatant x 1.9. The results of the HSAMMS-associated radioactivity were compared with a radiochromatography analysis performed on SG strips (ITLC-SG) with saline as the mobile phase at different times after the preparation for up to 94 h.

## 2.7. Biodistribution and *in vivo* stability of $^{90}$Y-labelled HSAMMS

The entire animal study conformed to ethical guidelines and complied with the European Council Directive (86/609/EEC) and the rules for animal care proposed by the Serbian Laboratory Animal Science Association (SLASA). The animals were kept in a well-ventilated, temperature-controlled (22±1 °C) animal room for a few days prior to the experimental period and provided with food and water. Fifteen healthy Wistar rats (100±10 g body weight, 4 weeks of age, Institute of Biology, Vinča Institute of Nuclear Sciences) were intravenously injected with 0.1 ml of the microsphere suspension containing 0.1 mg of $^{90}$Y-HSAMMS (approx. 2.5 MBq) through the tail vein. Five rats for each time point (1, 24, 72 h after injection) were sacrificed by cervical dislocation. The following organs were dissected out: muscle (hind leg), bone (the complete left femur), heart, lungs, liver, kidneys, spleen, stomach (emptied) and intestine (emptied) and a blood sample (1ml). Selected organs and tissues were disintegrated and solubilized in a final volume of 10 ml to reach an identical geometry and similar probe density for bremsstrahlung measurements of the $^{90}$Y radioactivity. The uptake into the organs was expressed as a percentage of the injected total radioactivity (%ID/organ) and per ml of blood, except for the radioactivity in the muscle, which was estimated by assuming a muscle weight of 40% of the total body weight. The amount of injected activity was calculated from the activity of the injection syringes before and after the injection because a significant quantity of the labelled microspheres stay attached to the plastics of the syringe. The results are reported as mean values ± standard deviation.

## 3. Results and discussion

### 3.1. Magnetite and ferrofluid preparation

In the absence of any surface coating, magnetic nanoparticles have hydrophobic surfaces with a large surface-area to volume ratio. Due to the hydrophobic interactions between the particles, they agglomerate and form large clusters with very strong magnetic dipole-dipole attractions among them [31]. In order to obtain a stable colloidal dispersion with the desired magnetic characteristics, different coating materials such as starch [32], polyethylene glycol/polyacril acid (PEG/PAA) [33], polylactic acid (PLA) [34], citric acid [35], oleic acid [36], decanoic acid and nonionic acid [37] have been applied. In the present study a citric acid coating was used to prevent the magnetite aggregation and to achieve good colloidal stability. The obtained citric acid-coated magnetite nanoparticles kept their colloidal characteristics for up to 3 months, with a very low level of sedimentation. It was assumed that citric acid can be adsorbed onto the surface of the magnetite nanoparticles by coordinating ≡Fe-OH sites via one or two of the carboxylate functionalities, depending on the steric necessity and the curvature of the surface. The carboxylic acid group on the citric acid exposed to the solvent stabilized the fluids based on the electrostatic repulsion between the particles [38, 39, 40]. At a higher citric acid concentration a tighter packed surface structure can be supposed, e.g., instead of two, only one carboxylate group of citric acid will be bound to the ≡Fe-OH sites, forming the surface of magnetite particles [41]. Therefore, the presence of available carboxylic groups provides a route to an extended bond formation with albumin. Also, individual and separated magnetic nanoparticles allowed better encapsulation in the HSA microspheres than the aggregates.

The representative FTIR measurements for citric-acid-coated magnetite nanoparticles and pure citric acid are shown in Fig. 1. The citric acid coating of the magnetite was confirmed by the presence of the peaks at 1560 and 1400 $cm^{-1}$. The 1695 $cm^{-1}$ peak, assignable to the C=O vibration (asymmetric stretching) from the COOH group in neat citric acid, after the binding of the citric acid to the magnetite surface, was shifted to a lower wave number. The neighbor band at 1400 $cm^{-1}$ was assigned to the symmetric stretching of C=O. The low-intensity peak at 540 $cm^{-1}$ can be associated with the stretching and torsion vibration modes of the magnetite. The large and intense band at approximately 3200 $cm^{-1}$ could be assigned to the structural OH groups as well as to the traces of molecular water and citric acid.

### 3.2. Preparation of the HSAMMS

The microspheres and particles designed for use as radionuclide carriers in radionuclide therapy must comply with the high standards relating to biocompatibility and biodegradability. From the production side, they must be easy to radiolabel and provide a high radiochemical stability of the labelled products against leaching of the radionuclide *in vivo* [42]. The high density of the $^{90}$Y-labelled glass microspheres used in the conventional therapy gave rise to problems during injection and, due to the non-biodegradability of the particles, they usually cause anatomical distress [43]. Resin microspheres are biocompatible but not biodegradable and suffer from a low specific activity, which requires the administration of large amounts of particles causing different adverse side-effects, like post-embolization syndrome [44]. To overcome these problems we synthesized biodegradable magnetic microspheres made from human serum albumin. Since the chemically cross-linked albumin microspheres showed non-uniform sizes and aggregates and due to the fact that heat-stabilized HSAMMS were found to be more stable [45], in our preparation we used the thermal denaturation method. Higher temperatures and longer heating times generally produce harder, less porous and more slowly degradable spheres [46]. Patil also reported that the stabilization of albumin microspheres (without loaded magnetite) at low temperatures (100–105 °C) resulted in a less smooth spherical geometry and some irregularity in the shape than when stabilizing at a higher temperature (170–175 °C) [46]. The reason for the smooth and compact surface of the albumin microspheres without magnetite is attributed to the preparation procedure in which the inner albumin phase of the water-in-oil emulsion was immediately solidified at high temperature with constant stirring. To investigate the effect of different temperatures on the geometry of the microspheres, the preparation of the HSAMMS was carried out at 100 °C and at higher temperatures (up to 170 °C). All the HSAMMS batches gave the same rough surface morphology. Increasing the temperature to 170 °C does not affect the surface morphology and a smooth surface was not formed, probably due to the presence of magnetite.

We found that several factors determined the final magnetite loading of the microspheres, such as the magnetite-to-albumin ratio, the stirring rate of the emulsion and the average size of the MS. For the synthesis of the HSAMMS we used the highest possible ratio of ferrofluid to albumin (obtained in our experiments) in order to have a better magnetic response. Besides the amount of magnetite material, its location and long retention inside the albumin microspheres are important factors when considering the characteristics of any magnetic microspheres. Even microspheres with adsorbed magnetic particles will respond to an external magnetic field, but they would not be considered as a stable as the adsorbed magnetic particles may detach from the surface after their introduction to a physiological medium.

### 3.3. Structure and thermal analysis of citric acid-coated magnetite and HSAMMS

The crystal structure of the pure $Fe_3O_4$ as well as $Fe_3O_4$ in both citric acid-coated magnetite and HSAMMS was checked using the X-ray diffraction technique (XRD). Fig. 2 shows the XRD pattern of the HSAMMS. All the reflections of the crystalline $Fe_3O_4$ phase are indexed in the spinel space group $Fd\bar{3}m$. No contamination by another phase was observed. The broad background is characteristic of amorphous HSA. The mean crystallite diameter of 11 nm was estimated for the $Fe_3O_4$ using Scherer's equation and the peak half-height width of the (311) reflections. The same values were obtained for pure $Fe_3O_4$ and $Fe_3O_4$ in the citric acid-coated magnetite.

Fig. 3 shows the TGA curves of the citric acid-coated magnetite and the HSAMMS (the corresponding derivatives are not shown). The thermogravimetric data showed that in the case of the coated NPs an initial weight loss of 6.1% (up to 127 °C) is due to the evaporation of physically adsorbed water. In the next step, up to 159 °C, the weight loss is 1.3%. The main step along the decomposition process was observed in the interval 159–341 °C, with a weight loss of 42%. No further weight loss was observed above 341 °C, with a mass residue of 49%. The product of the thermal degradation was some iron-oxide phase and some carbon-based residuals.

The thermogravimetric curve of the HSAMMS showed that the thermal degradation occurs in three steps. A first step, related to dehydration, was observed up to 130 ºC, with a loss of weight of 9.8%. For the second (200–383 °C) and third (up to 580 °C) steps the observed weight loss was 33 and 51%, respectively. The final mass residue was 6% of the initial material, suggesting that the products of the thermal degradation were mainly $Fe_3O_4$ and residual carbon-related compounds.

**3.4. FESEM, SEM/FIB and HRTEM analysis of magnetite, citric acid-coated magnetite and HSAMMS**

The morphology and internal composition of the HSAMMS were assessed by FESEM and dual-beam SEM/FIB microscopy. As can be seen from the secondary-electron images shown in Figures 4a and 4b the HSAMMS particles are spherically shaped with sizes spanning from 10 to 70 μm. The surface of the microspheres is rough and, according to an energy-dispersive X-ray analysis (EDX), no detectable amounts of Fe-containing material can be found on their surface. The internal 3D distribution of the nanoparticles was analyzed by cross-sectioning a single microsphere using FIB, and further reconstruction of the slices into a 3D perspective (Figure 4 c, d). We found a cluster distribution of high-contrast areas corresponding to the magnetic nanoparticles inside the microsphere, as confirmed by an EDX analysis of these spots (see Fig. 5).

Close to circumference some of the smaller spheres were electron transparent to the electron beam. Fig. 6 shows bright-field TEM images of the magnetite nanoparticles embedded in the organic matrix. It is mainly spherical particles that are seen, with a small amount of embedded aggregates. The magnetic nanoparticles appeared to be uniform in size from TEM images, with an average size of about 8–10 nm.

### 3.5. Hysteresis loops

The hysteresis loops of the samples M(H), recorded at 300 K, are shown in Fig. 7. The absence of any magnetization saturation was observed in all the samples. The M(H) curves showed coercivity ($H_C$) and remanence ($M_r$) values close to zero, indicating that the particles are superparamagnetic for a time scale of the order of seconds. The saturation magnetization $M_S$ value was estimated by extrapolation of the $M$ vs. $1/H$ curve when $1/H \to 0$. We found that the sample $Fe_3O_4$ has $M_S = 61.5$ emu/g, while $Fe_3O_4$/CA and HSAMMS have $M_S = 25$ emu/g and $M_S = 2$ emu/g, respectively. The values for the two last samples are the result of the presence of the organics, citric acid (CA) and citric acid with human serum albumin, and the reducing effective mass of magnetite. The fact that the HSAMMS are magnetic ensures that the engendered material, as an effective radionuclides carrier, can be moved under a moderate magnetic field strength.

### 3.6. Labelling of HSAMMS with yttrium-90

The radio-tracer technique, possessing high sensitivity, credibility and freedom from interference, is a relevant approach to study the biodistribution of novel particles [47]. Nevertheless, there are still some limitations to the methods that can be used for radiolabelling [48]. Firstly, the binding of the radionuclide has to be irreversible in order to prevent their escape to other tissues or organs. This problem exists even in newly developed, DOTA (1,4,7,10-tetraazacyclododecane-N, N',N'',N'''-tetraacetic acid) derived HSA microspheres, since the radiation induces the cleavage of the thiourea linker of the $^{90}$Y-DOTA-HSAMMS and makes it unsuitable for therapeutic applications [16]. Secondly, due to the necessarily short half-life time of $^{90}$Y, the manufacturing process is time limited and too complex because of the risk of contamination. The labelling of the HSAMMS was achieved by adding a YCl$_3$ solution to a slightly acidic microsphere water suspension and subsequent shaking for 15 min at 80 °C. The maximum labelling yield of $^{90}$Y-HSAMMS, achieved under the experimental conditions previously described, was 67%, determined by ITLC-SG. A similar result (71%) for the labelling yield was obtained by magnetic precipitation. For the following biodistribution studies the main step was to remove the excess of unbound $^{90}$Y, as free $^{90}$Y present in the suspension with labelled products may accumulate in the bone. Therefore, the $^{90}$Y-HSAMMS were retrieved using magnetic decantation, and the supernatant with unbound $^{90}$Y was withdrawn and the labelled HSAMMS were additionally washed several times with deionizied water, so the radioactivity that is present in the bone will be an indicator of the *in vivo* instability of the $^{90}$Y-HSAMMS.

### 3.7. *In vitro* stability studies

The results of both methods (magnetic precipitation and ITLC-SG) showed that less than 4.1% of the $^{90}$Y dissociated from the surface of the $^{90}$Y-HSAMMS in saline, and a slightly increased percentage (6.9%) in the serum, after 1 h (Fig. 8). Even without the surface modification the $^{90}$Y remained bound to the HSAMMS for a time period sufficient for a further *in vivo* application. The non-particle bound radioactivity consists only of the free $^{90}$Y, moving with the solvent front ($R_f$=0.9-1.0), and no other degradation products were found. The scanning electron microscopy images (not shown) revealed that the microspheres have the same characteristic morphology for a 7-day incubation period in HSA at 37 °C. Therefore, the leaching of the $^{90}$Y is only the result of its detaching from the HSAMMS surface, but not the degradation of the HSAMMS.

### 3.8. Biodistribution study

The biodistribution and *in vivo* stability of the labelled HSAMMS were studied, after the intravenous injection of approximately 2.5 MBq of $^{90}$Y-HSAMMS into the tail veins of rats, up to 72 h after the application.

The $^{90}$Y-HSAMMS were completely trapped in the capillary bed of the lungs, which served in our experiments as a model for the vascular system of the solid tumors [49]. The high uptake (84.81%ID) of the $^{90}$Y-HSAMMS 1h after injection (Fig. 9) is expected since the HSA microsphere action mechanism is based mainly on the particle size. It was reported that HSA microspheres in the size range 5–30 μm were accumulated almost entirely in the lung in the first pass of circulation through the pulmonary artery following intravenous administration [50- 52].

The disappearance of the radioactivity from the lungs was used as a measure of the *in vivo* stability of the microspheres and the attached $^{90}$Y. A significantly high percentage of $^{90}$Y-HSAMMS was present in the lung after 72 h (82.67%ID), showing a satisfactory *in vivo* stability of the $^{90}$Y-HSAMMS. Since the labelled HSAMMS retained in the lung for at least one half-life, the release of the $^{90}$Y is proved by the low radioactivity present in the bone, and consequently the damage to the surrounding healthy tissue is negligible. Both the high *in vivo* retention of radioactivity in the lung as well as the *in vitro* stability studies are proof that the $^{90}$Y-HSAMMS stay intact in the lung for at least 72h, giving the possibility to use them in a combined, dual radionuclide-hyperthermia therapy. It is reported that the *in vivo* application of hyperthermia showed increased perfusion in the tumor region, leading to a higher oxygen concentration [53]. Most human tumors increased the blood flow under hyperthermia and for hours later. Higher perfusions can increase the radionuclide delivery and oxygenation required to enhance the free-radical-dependent cell death in radiotherapy [54].

The results of the biodistribution of the $^{90}$Y-HSAMMS 1h after injection (Fig. 9) showed that a minor percentage of activity was found in other organs: 4.95 %ID, liver; 4.49% ID, intestines; 1.03%ID, kidney and 0.21 %ID, spleen. The tissue distribution of the particulate materials depends on the size of the particles injected. Particles larger than 10 μm should be located in the lung, particles over 100 nm accumulate preferentially in the liver and spleen (0.2–3 μm), while smaller particles (<30 nm) tend to accumulate in a relatively higher concentration in the bone marrow [55, 56]. The dominant tissue accumulation of $^{90}$Y-HSAMMS in one organ, the lung, is in agreement with the FESEM analysis of the size distribution of the HSAMMS, showing the majority of the particles are larger than 10 μm. The above biodistribution screening method in rats was used to evaluate the radiolabelled magnetite particles for possible application in the local radionuclide therapy of different tumors. Owing to its size, the $^{90}$Y-HSAMMS could be used in the SIRT of hepatic metastases by intra-arterial administration.

Finally, the heating efficiency of the HSAMMS obtained from the power-absorption measurements ($f$=357 kHz, H = 47 kA/m) showed a heating rate of 3.95 K/min, which could be enough to reach the temperature range required for thermally induced cell apoptosis by hyperthermia. The SPA value obtained from these measurements was 13.8 W/g. Although this is a moderate value when compared to previous reports [6,7], the heating rate could be further increased by increasing the actual concentration of the magnetic microsphere in the target area.

## 4. Conclusions

We have synthesized a magnetic, microstructured composite based on citric acid-coated $Fe_3O_4$ nanoparticles encapsulated into biodegradable HSA microspheres. Without any further chemical surface modification, *in vivo* stable $^{90}$Y-labelled HSAMMS were prepared. The created $^{90}$Y-HSAMMS should be able to deliver high doses of radiation to the target area. The specific-absorption-rate measurements showed that radiolabelled HSAMMS are promising microsystems for use in a combined cancer treatment based on radionuclide radiation and hyperthermia, profiting from the synergy of both protocols. The reported procedure for engineering magnetic microspheres can be used for the synthesis of suitable composite materials for cancer therapy as well as for drug/radionuclide-delivery magnetic materials.

**Acknowledgments**

The authors acknowledge the financial support from the Spanish Ministry of Education and the Serbian Ministry of Education and Science through the Serbian-Spanish bilateral project AIB2010SE-00202. The Serbian Ministry of Education and Science supported this work financially through the projects Grant No. III45015. The Spanish Ministerio de Ciencia e Innovacion supported this work through Project MAT2010-19326. We thank Dr. Jelena Rogan for performing the TGA measurements.

**Figure caption**

Figure 1. FTIR spectra of citric acid-coated magnetite ($Fe_3O_4$/CA) and citric acid (CA)

Figure 2. X-ray diffraction pattern of HSAMMS. The arrows indicate the indexing of the peaks corresponding to the spinel structure of $Fe_3O_4$

Figure 3. TGA curves for citric acid-coated magnetite ($Fe_3O_4$/CA) and HSAMMS ($Fe_3O_4$/CA/HSA)

Figure 4. Upper panels (a,b): FESEM images of HSAMMS. Lower panels (c,d): 3D reconstruction of 50 cross-sectional slices on a single microsphere, showing the distribution of the magnetic nanoparticles (blue spots) within the microspheres

Figure 5. Upper panels: Compositional mapping from EDX analysis of a selected area (pink boxes) on a cross-sectioned HSAMMS showing the iron contents (green colour). Lower panel: the corresponding EDX spectrum over the same selected area. The small Ga content comes from the ion beam

Figure 6. HRTEM pictures of $Fe_3O_4$/CA (a, b) and selected area of HSAMMS (c)

Figure 7. Magnetization versus field at room temperature. Inset: Hysteresis loop for the HSAMMS

Figure 8. *In vitro* stability of $^{90}$Y-HSAMMS in saline and human serum up to 96 h

Figure 9. Biodistribution of $^{90}$Y-HSAMMS in different tissues after 1, 24, 72 h of intravenous administration in normal Wistar rats. The results are expressed as % ID/organ (mean five rats ± standard deviations). The inset: results for tissues without lungs

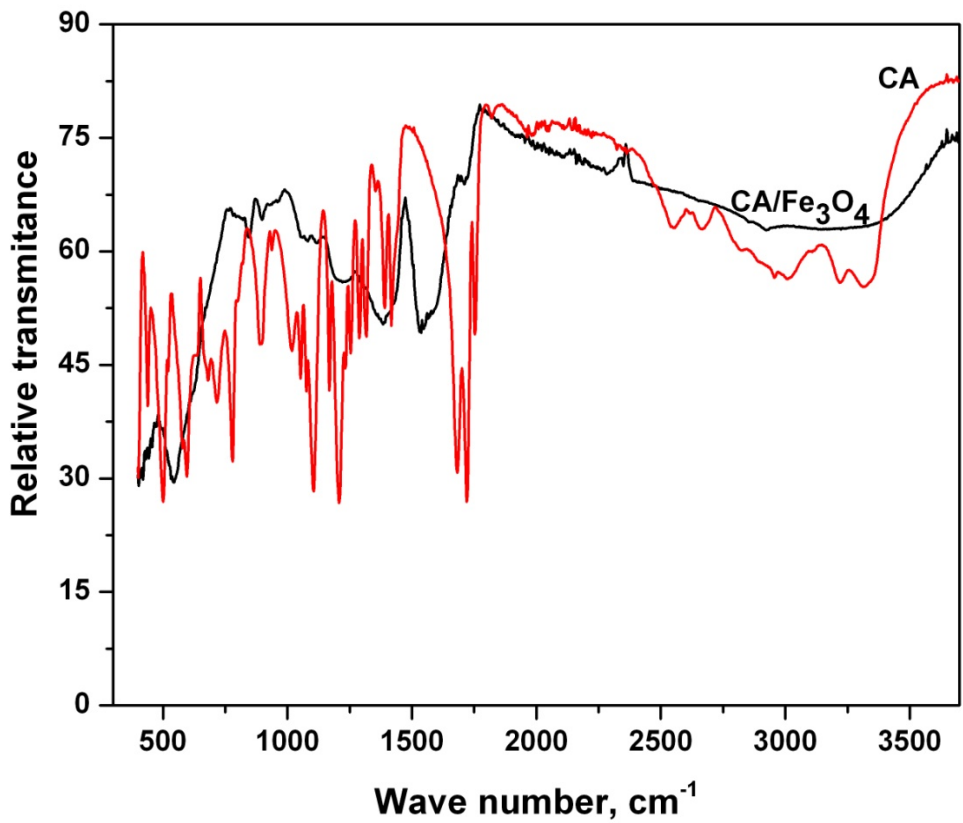

*Figure 1.*

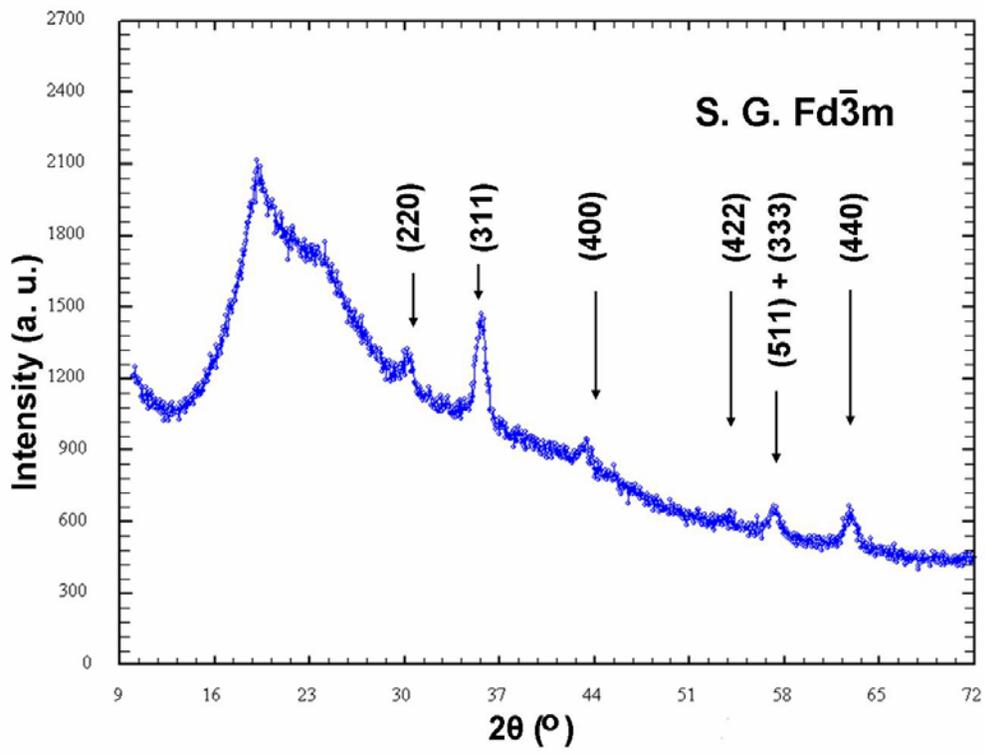

*Figure 2.*

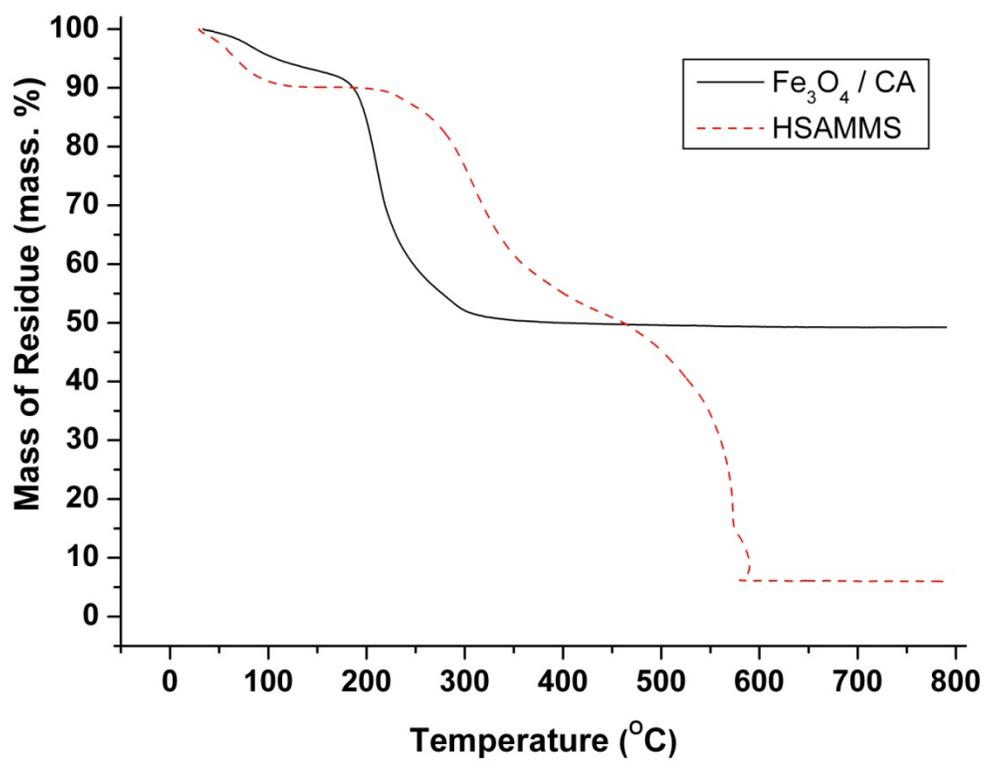

*Figure 3.*

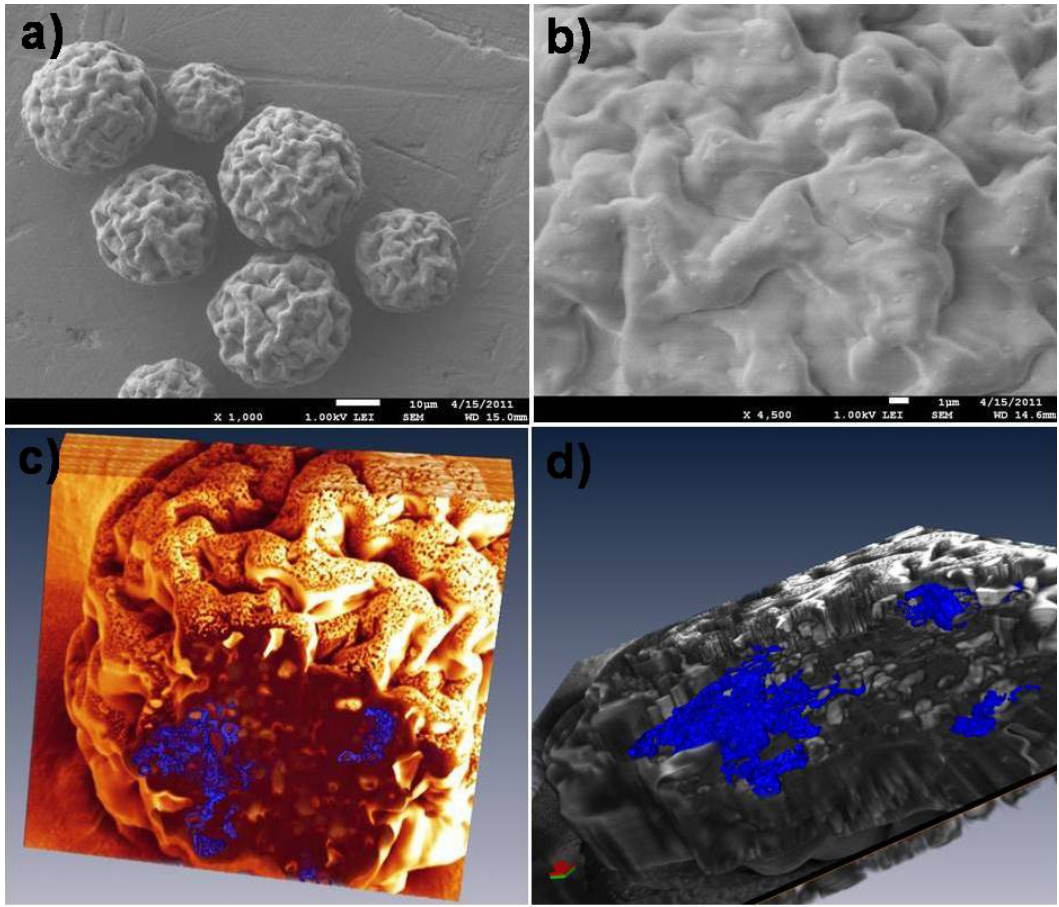

*Figure 4.*

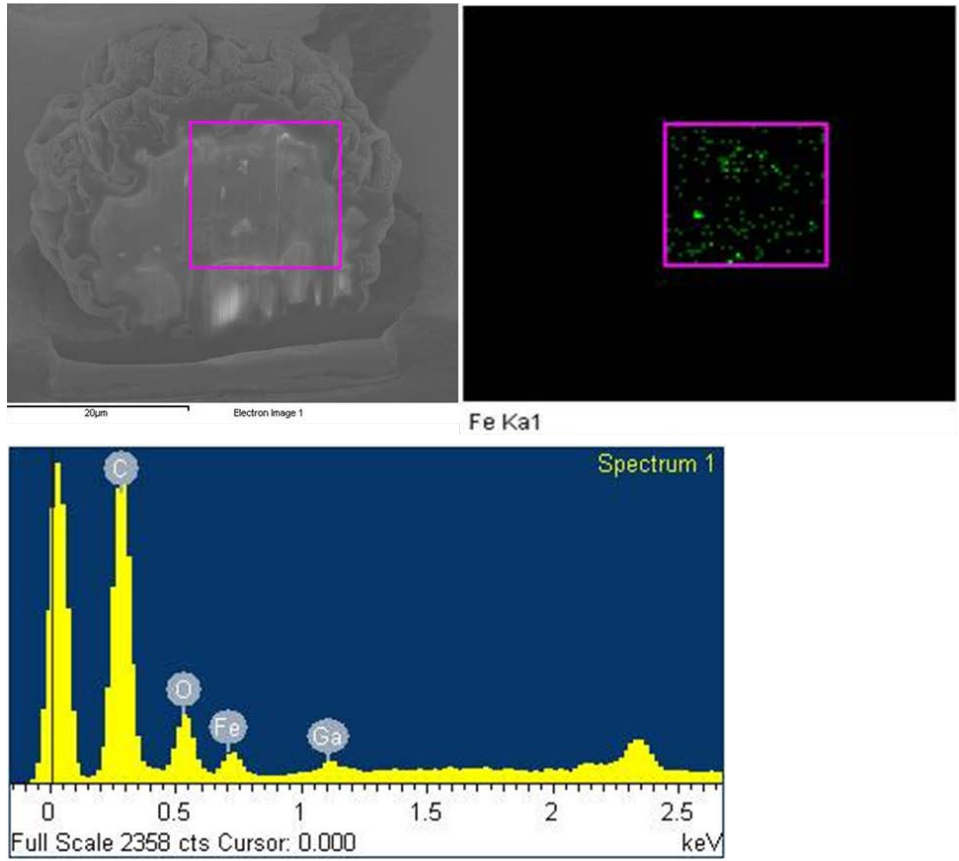

*Figure 5.*

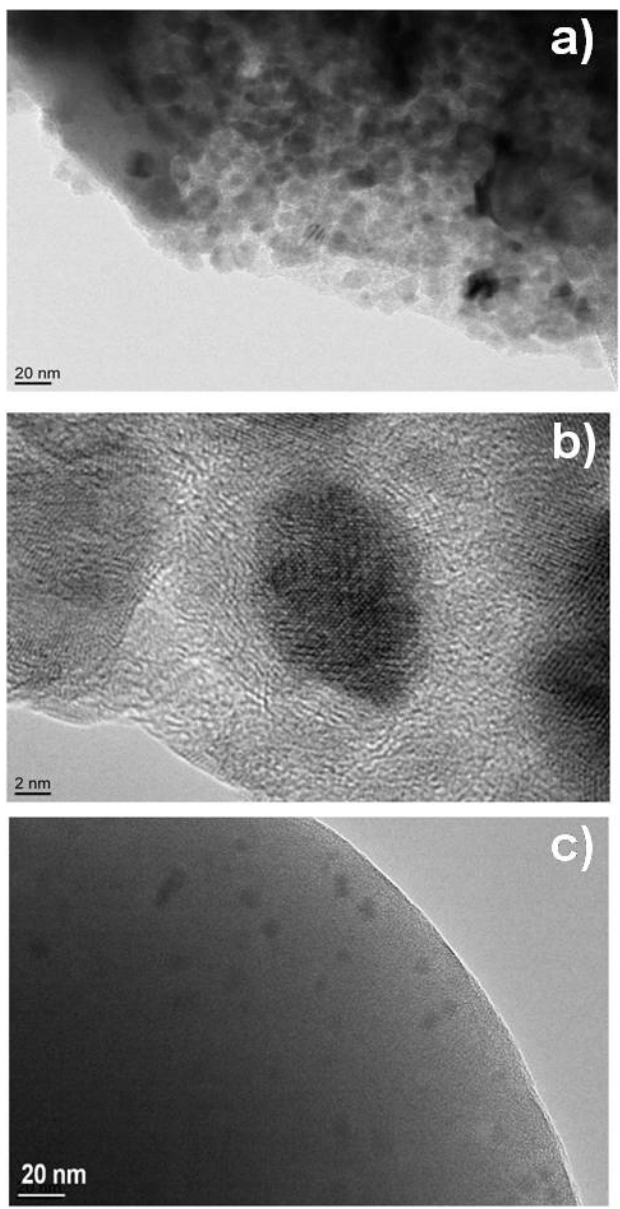

*Figure 6.*

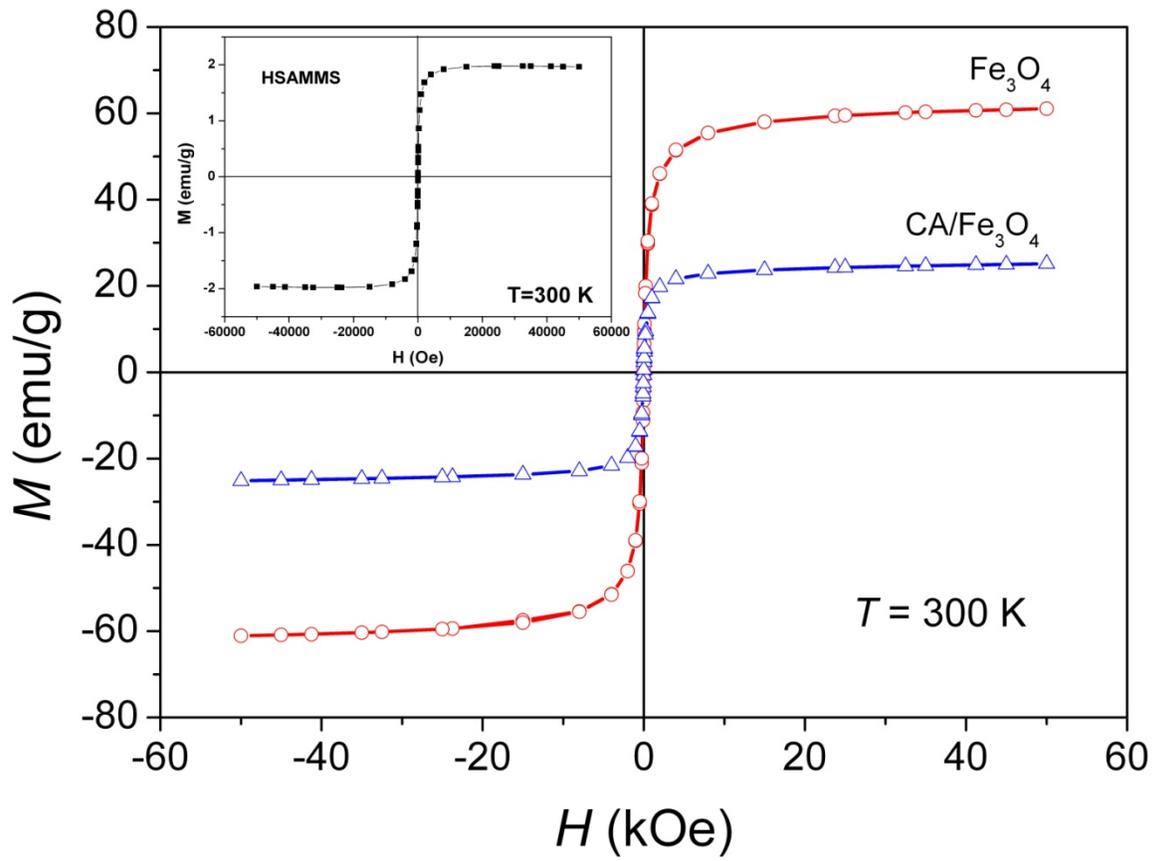

*Figure 7.*

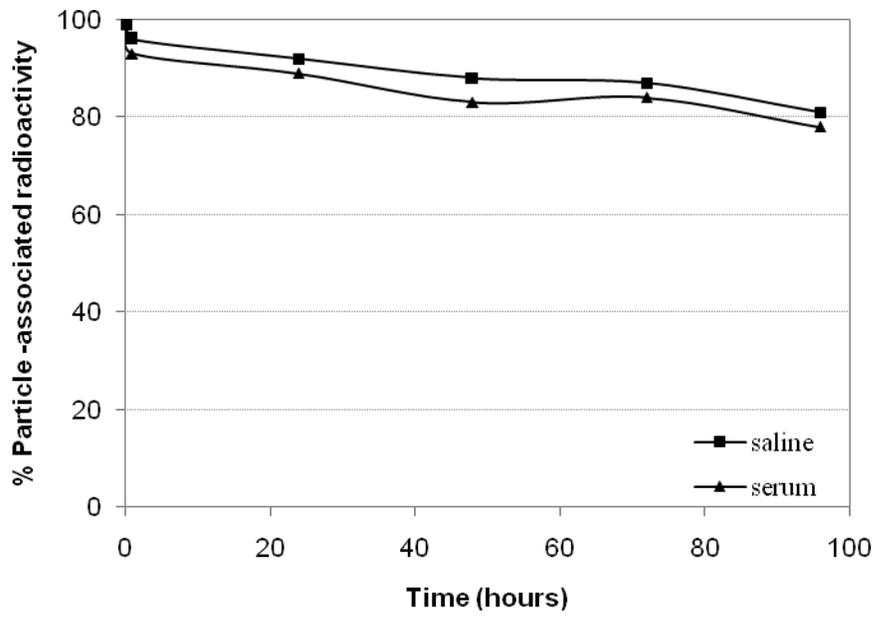

*Figure 8.*

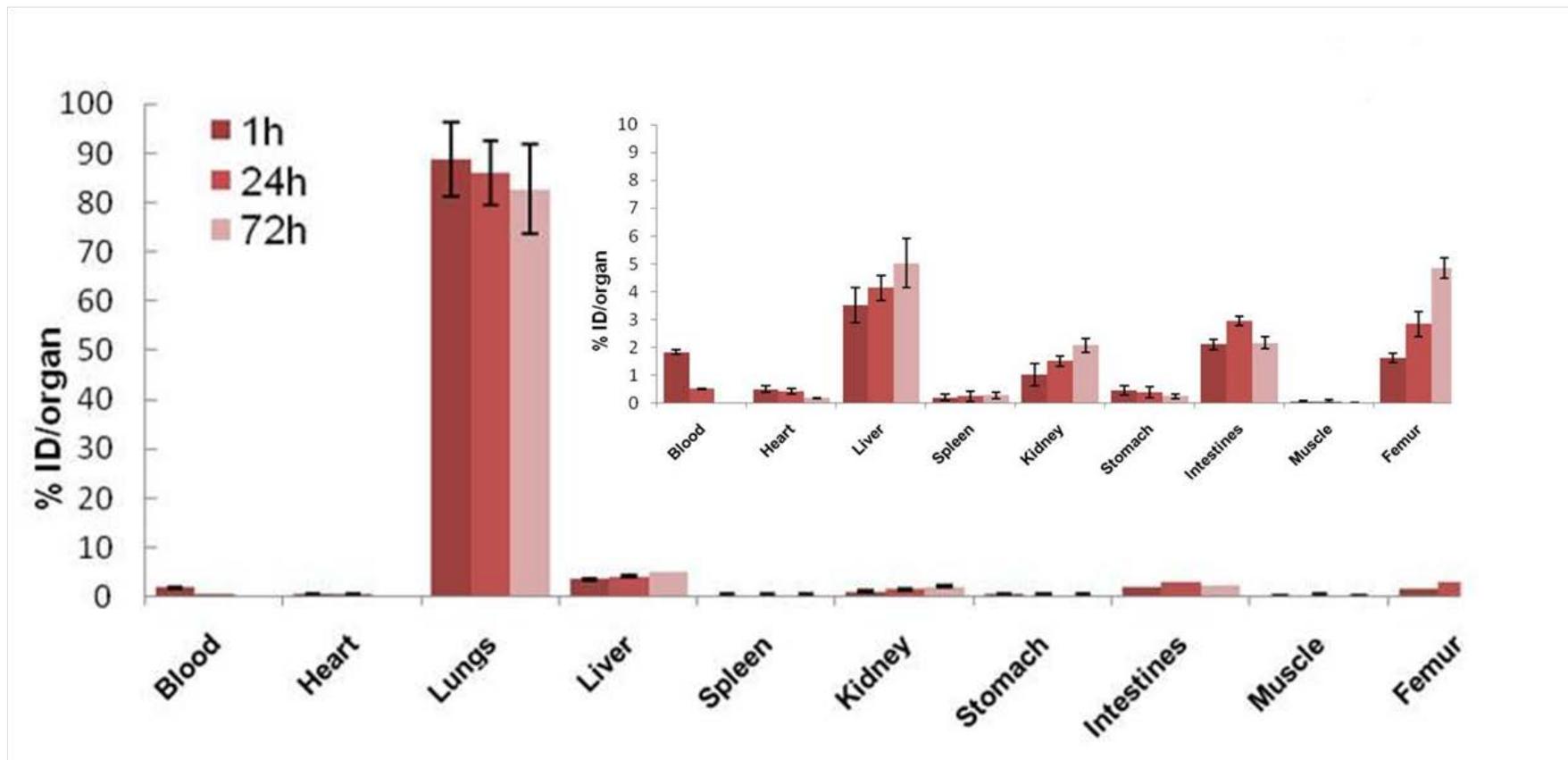

*Figure 9.*